\documentclass[12pt]{article}			
\usepackage{epsfig}

\makeatletter			
\@addtoreset{equation}{section}
\def\theequation{\thesection.\arabic{equation}}
\makeatother			

\newcommand{\be}{\begin{equation}}
\newcommand{\ee}{\end{equation}}
\newcommand{\bea}{\begin{eqnarray}}
\newcommand{\eea}{\end{eqnarray}}
\newcommand{\smallz}{{\scriptscriptstyle Z}} 
\newcommand{\smallw}{{\scriptscriptstyle W}} %
\newcommand{\mz}{M_\smallz}
\newcommand{\mw}{M_\smallw}

\newcommand{\fr}{\frac}

\def \gev  {\mbox{ GeV}}
\def \mev  {\mbox{ MeV}}

\def \psl  {p \kern-.45em{/}}
\def \qsl  {q \kern-.45em{/}}
\def \qslov {\overline{q \kern-.45em{/}}}
\def \pslov {\overline{p \kern-.45em{/}}}

\def \lsim {\raisebox{-.7ex}{$\stackrel{\textstyle <}{\sim}\,$}}
\def \gsim {\raisebox{-.7ex}{$\stackrel{\textstyle >}{\sim}\,$}}

\def \xiw  {\xi_{\mbox{\footnotesize{w}}}}
\def \xiz  {\xi_{\mbox{\footnotesize{z}}}}
\def \xig  {\xi_\gamma}
\def \sov  {\overline{s}}
\def \mov  {\overline{m}}

\def \Aov  {\bar{A}}
\def \LM   {\ln \! \left(\frac{M^2-s}{M^2} \right)}
\def \LMxiw {\ln \!\left(\frac{M^2\xiw-s}{M^2\xiw} \right)}
\def \LS   {\ln \!\left(\frac{\sov-s}{\sov} \right)}
\def \Agamma {A^{\gamma}}

\begin{document}

\begin{titlepage}
\begin{flushright}
        \small
        NYU-TH/98-04-01\\
        hep-ph/9804309 \\
	April 1998
\end{flushright}

\begin{center}
\vspace{0.5cm}
{\LARGE \bf Radiative Corrections to \boldmath{$W$} and Quark \\
           Propagators in the Resonance Region}

\vspace{1cm}
\renewcommand{\thefootnote}{\fnsymbol{footnote}}
{\bf    M.~Passera\footnote
                {E-mail address: massimo.passera@nyu.edu} and
        A.~Sirlin\footnote
                {E-mail address: alberto.sirlin@nyu.edu}}
\setcounter{footnote}{0}
\vspace{.8cm}

{\it    Department of Physics, New York University, \\
        4 Washington Place, New York, NY 10003, USA}

\vspace{1.8cm}

{\large\bf Abstract}
\end{center}

\noindent
We discuss radiative corrections to $W$ and quark propagators in 
the resonance region, $|s-M^2|$ \lsim $M \Gamma$. We show that conventional 
mass renormalization, when applied to photonic or gluonic corrections,
leads in next to leading order (NLO) to contributions proportional to 
$[M\Gamma/(s-M^2)]^n$, $(n=1,2\ldots)$, i.e. to a non-convergent series 
in the resonance region, a difficulty that affects all unstable particles 
coupled to massless quanta. A solution of this problem, based on the 
concepts of pole mass and width, is presented.
It elucidates the issue of renormalization of amplitudes involving
unstable particles, and automatically circumvents the problem of 
apparent on-shell singularities. The roles of the 
Fried-Yennie gauge and the Pinch Technique prescription are discussed.
Because of special properties of the photonic and gluonic contributions, 
and in contrast with the $Z$ case, the gauge dependence of the 
conventional on-shell definition of mass is unbounded in NLO.
The evaluations of the width in the conventional and pole formulations 
are compared and shown to agree in NLO but not beyond.


\end{titlepage}

\section{Introduction}

The aim of this paper is to study the radiative corrections to 
$W$ and unstable fermion propagators in the resonance region.
Calling $s$ the invariant momentum transfer, this is the region 
$|s-M^2|$ \lsim $M \Gamma$, where $M$ and $\Gamma$ are the mass and the 
width of the unstable particles. The $W$ analysis is a natural 
counterpart of the study of the $Z^0$ propagator that has played a 
major role in the 
interpretation of electroweak physics in the resonance region. 
For some time it has been known that the conventional on-shell definition 
of mass, 
\be
	M^2 = M_0^2 + Re A(M^2), 
\label{eq:M}
\ee
where $M_0$ is the unrenormalized mass 
and $A(s)$ is the transverse boson self-energy (including tadpole 
contributions), is gauge-dependent in $O(g^4)$ and higher 
\cite{Si91a,Si91b,PaSi96}.
In the $Z^0$ case, the gauge dependence of $M$ is $\lsim 2\mev$ in $O(g^4)$
but becomes unbounded in  $O(g^6)$ \cite{PaSi96}.
On the other hand, the complex-pole 
position 
\be
        \sov = m_2^2 -i m_2 \Gamma_2 = M_0^2 + A(\sov)
\label{eq:sov}
\ee
is gauge-invariant. Thus, a gauge-invariant definition can be achieved 
by identifying the mass with $m_2$ or appropriate combinations of $m_2$
and $\Gamma_2$. In particular, it has been shown \cite{Si91a}
that the $Z$ mass 
measured at LEP can be identified with 
\be
        m_1 = \left(m_2^2 +\Gamma_2^2\right)^{1/2}.
\label{eq:m1}
\ee 
In Eqs.~(\ref{eq:sov},~\ref{eq:m1}) we have followed the notation 
introduced in Eqs.~(4) and (15) of Ref. \cite{Si91a}.

In the $W$ case one expects similar theoretical features. However, 
as shown in Section 2, a new problem emerges: in the treatment of the 
photonic corrections the conventional mass-renormalization procedure
generates
contributions proportional to $[M\Gamma/(s-M^2)]^l$, $(l=1,2\ldots)$, in
next to leading order (NLO). Thus, one obtains an expansion that does not 
converge in the resonance region! These theoretical features are generally 
present whenever the unstable particle is coupled to massless quanta. 
In Section 2 we present a solution of this problem based on the 
concepts of pole mass and width.
It automatically circumvents the problem of apparent on-shell singularities
and, more generally, it elucidates the issue of renormalization 
of amplitudes involving unstable particles. 
The roles of the Fried-Yennie gauge and the Pinch Technique are discussed
in Section 3. In contrast to the $Z$ case, we show 
that, because of special features of the bosonic and gluonic contributions, 
the gauge dependence of the conventional on-shell definition of mass is   
unbounded in NLO.
Section 4 discusses the overall corrections to the $W$ propagator in NLO.
In Section 5 the modified and conventional formulations of the $W$
width are compared and shown to agree in NLO, but not beyond.
Potential problems of the conventional definition of width emerging in
high orders of perturbation theory are discussed.
As a further illustration, in Section 6 we discuss the QCD corrections
to an unstable quark propagator in the resonance region.

\section{Photonic Corrections to the \boldmath{$W$} Propagator in 
the Resonance Region}

In order to illustrate the difficulties emerging in the  resonance region 
when the conventional mass renormalization is employed, we consider the 
contributions of the transverse part of the $W$ propagator in the loop of
Fig.1, with $l$ self-energy insertions. Writing the transverse $W$ 
self-energy in the form
\be
        \Pi^{\scriptscriptstyle (T)}_{\mu \nu}(q) = t_{\mu \nu}(q) A(s),
\ee
where $s\equiv q^2$ and  
$t_{\mu \nu}(q)=$ $g_{\mu \nu} - q_\mu q_\nu/ q^2$,
the contribution $A_{\smallw \gamma}^{(l)}(s)$ from Fig.1 to $A(s)$ is given 
by 
\be
        A_{\smallw \gamma}^{(l)}(s) = i e^2(\mu)
	\,\frac{t_{\mu \nu}(q)}{(n-1)}\,
	\mu^{4-n}
        \int \frac{d^n \!k}{(2\pi)^n} {\cal{D}}
		^{{\scriptscriptstyle (} \gamma {\scriptscriptstyle )}}
		_{\rho \beta} (k)
        {\cal{D}}^{\scriptscriptstyle (W,T)}_{\lambda \alpha}(p)
        {\cal{V}}^{\rho \lambda \nu} {\cal{V}}^{\beta \alpha \mu} 
        \left[\frac{A^{{\scriptscriptstyle (} s {\scriptscriptstyle )}}(p^2)}
                {p^2 -M^2 +i \epsilon} \right]^l,  
\label{eq:A-l-insertions}
\ee
where $p=q+k$ is the $W$ loop-momentum,
\be
        {\cal D}^{{\scriptscriptstyle (} \gamma {\scriptscriptstyle)}}
		_{\rho \beta} (k) = -\frac{i}{k^2}
        \left(g_{\rho \beta} + (\xig -1)
        \frac{k_\rho k_\beta}{k^2} \right),
\ee
\be
        {\cal D}^{\scriptscriptstyle (W,T)}_{\lambda \alpha}(p)=
        \frac{-i}{p^2 -M^2 +i\epsilon} 
        \left(g_{\alpha \lambda} - \frac{p_\alpha p_\lambda}{p^2} \right),
\label{eq:Wprop-landau}
\ee
\be
        {\cal V}^{\beta \alpha \mu} = (2p-k)^\beta g^{\alpha \mu} +
        (2k-p)^\alpha g^{\beta \mu} -(k+p)^\mu g^{\beta \alpha},
\label{eq:vertex}
\ee
$\xig$ is the photon gauge parameter and 
$A^{{\scriptscriptstyle (} s {\scriptscriptstyle )}}(p^2)$ stands for 
the $W$ transverse self-energy with the conventional mass renormalization
subtraction:
\bea
	A^{{\scriptscriptstyle (} s {\scriptscriptstyle )}}(p^2) &=& 
			\mbox{Re}\left(A(p^2)-A(M^2) 
			\right) + i \mbox{Im}A(p^2)  \nonumber  \\
		&=& A(p^2) - A(M^2) + i\mbox{Im}A(M^2).
\eea
We recall that, in leading order, $i\mbox{Im}A(M^2)=-iM\Gamma$.
Eq.~(\ref{eq:Wprop-landau}) corresponds to the choice $\xiw=0$ for the 
$W$ gauge parameter $\xiw$ (Landau gauge). We note that each
insertion of 
$A^{{\scriptscriptstyle (} s {\scriptscriptstyle )}}(p^2)$ 
is accompanied by an additional denominator
$[p^2-M^2 +i\epsilon]$. Thus, Eq.~(\ref{eq:A-l-insertions}) may be 
regarded as the $l$th term in an expansion in powers of 
$$
        \left[A(p^2) - A(M^2) + i\mbox{Im}A(M^2) \right] 
        \left(p^2-M^2 +i\epsilon\right)^{-1}. 
$$
As $A(p^2) - A(M^2) = O[g^2(p^2-M^2)]$ for $p^2 \approx M^2$, the 
contribution 
$[A(p^2) - A(M^2)]\times$ $(p^2-M^2 +i\epsilon)^{-1}$ is of $O(g^2)$ 
throughout the 
region of integration. Therefore, each successive insertion leads to 
corrections of higher order in $g^2$. However, as 
$i\mbox{Im}A(M^2) \approx -iM\Gamma$ is not subtracted, the combination
$i\mbox{Im}A(M^2)/(p^2-M^2 +i\epsilon)$ may lead to terms of $O(1)$
if the domain of integration $|p^2-M^2|$ \lsim $M \Gamma$ is important.
In fact, the contribution of $[i\mbox{Im}A(M^2)/(p^2-M^2 +i\epsilon)]^l$
to Eq.~(\ref{eq:A-l-insertions}) is, to leading order,
\be
        A_{\smallw \gamma}^{(l)}(s) =\frac{(-iM\Gamma)^l}{l!}
        \frac{d^l}{d(M^2)^l} A_{\smallw \gamma}^{(0)}(s) + \ldots
\label{eq:A-l-insertions-Im}
\ee
where $A_{\smallw \gamma}^{(0)}(s)$ represents the diagram without 
self-energy insertions and the dots indicate additional contributions
not relevant to our argument.

In the resonance region the inverse zeroth order propagator
is proportional to $(s-M^2+iM\Gamma) =O(g^2)$. Therefore, in NLO, 
contributions of $O[\alpha(s-M^2)]$ are retained but those of 
$O[\alpha(s-M^2)^2]$ are neglected. Explicit evaluation of 
$A_{\smallw \gamma}^{(0)}(s)$ in NLO leads to 
\be
        A_{\smallw \gamma}^{(0)}(s) = \frac{\alpha}{2\pi}
        \left[(\xig-3)(s-M^2)\LM + \ldots\right].
\label{eq:A-0-insertions-partial}
\ee
Inserting Eq.~(\ref{eq:A-0-insertions-partial}) into 
Eq.~(\ref{eq:A-l-insertions-Im}) we obtain 
\bea
	A_{\smallw \gamma}^{(1)}(s) &=& \frac{\alpha}{2\pi}(\xig -3)
		\left( iM \Gamma \right) \left[ \LM + \fr{s}{M^2}
		\right] + \ldots,
			          \nonumber \\	
        A_{\smallw \gamma}^{(l)}(s) &=&\frac{\alpha}{2\pi}(\xig -3)
        \frac{(s-M^2)}{l(l-1)} 
        \left(\frac{-iM\Gamma\phantom{^2}}{s-M^2} \right)^{\!l} + \ldots, 
        \quad   (l \geq 2).
\label{eq:A-l-insertions-Im-again}
\eea
We see from 
Eq.~(\ref{eq:A-l-insertions-Im-again}) that Fig.1, evaluated with 
conventional mass renormalization, leads in NLO to a series in powers of 
$M\Gamma/(s-M^2)$, which does not converge in the resonance region.
Thus, rather than generating contributions of higher order in $g^2$,
each successive self-energy insertion gives rise to a factor  
$-iM\Gamma/(s-M^2)$, which is nominally of $O(1)$ in the resonance region 
and furthermore diverges at $s=M^2$!

Formally, the series $\sum^{\infty}_{l=0}A_{\smallw \gamma}^{(l)}(s)$
with $ A_{\smallw \gamma}^{(l)}(s)$ given by 
Eq.~(\ref{eq:A-l-insertions-Im}), can be resummed. In fact, it leads to 
\be
        \sum^{\infty}_{l=0} A_{\smallw \gamma}^{(l)}(s, M^2) = 
        A_{\smallw \gamma}^{(0)}(s, M^2-iM\Gamma) + \ldots.
\ee
Thus,
\be
 	\sum^{\infty}_{l=0} A_{\smallw \gamma}^{(l)}(s)=\frac{\alpha}{2\pi}
        \left[(\xig-3)(s-M^2+iM\Gamma)\ln
        \left(\frac{M^2-iM\Gamma -s}{M^2-iM\Gamma}\right) + \ldots\right].
\label{eq:A-0-insertions-partial-again}
\ee
Even if one accepts these ``a posteriori'' formal resummations, the 
theoretical situation in the framework of conventional mass 
renormalization is unsatisfactory.
In fact, in the conventional formalism, the $W$ propagator is inversely 
proportional to 
\be
        {\cal D}^{-1}(s) = s-M^2 +i M\Gamma 
		-\left( A(s)-A(M^2) \right) 
		-i M\Gamma \,\mbox{Re} A^\prime(M^2)
\label{eq:inverse_prop}
\ee
where $\Gamma$ is the radiatively corrected width and we have employed
its conventional definition
\be
	M \Gamma = -\mbox{Im}A(M^2)/[1-\mbox{Re}A'(M^2)].
\label{eq:usualwidth}
\ee
The contribution of Eq.~(\ref{eq:A-0-insertions-partial-again}) to 
${\cal D}^{-1}(s)$ is 
$$
        -\frac{\alpha}{2\pi} (\xig-3) \left[(s-M^2+iM\Gamma)\ln
        \left(\frac{M^2-iM\Gamma -s}{M^2-iM\Gamma}\right)
        +iM\Gamma \left(1+i\frac{\pi}{2} \right) \right] + \ldots
$$

We note that the last term is a gauge-dependent contribution not 
proportional to the zeroth order term $s-M^2+iM\Gamma$. 
As a consequence, in NLO the pole position is 
${\widetilde{M}}^2 -i\widetilde{M} \widetilde{\Gamma}$, where
\bea
	{\widetilde{M}}^2 &=& M^2[1-(\alpha/4)(\xig-3)(\Gamma/M)],
\label{eq:Mtilde}				\\
	\widetilde{\Gamma} &=& \Gamma [1-(\alpha/2\pi)(\xig-3)].
\label{eq:Gammatilde}	
\eea
As the pole position is gauge-invariant, so must be $\widetilde{M}$ and
$\widetilde{\Gamma}$. Furthermore, in terms of $\widetilde{M}$ and
$\widetilde{\Gamma}$, ${\cal D}^{-1}(s)$ retains the Breit--Wigner
structure. Thus, in a resonance experiment $\widetilde{M}$ and
$\widetilde{\Gamma}$ would be identified with the mass and width of $W$.
The relation $\widetilde{\Gamma} = \Gamma [1-(\alpha/2\pi)(\xig-3)]$
leads then to a contradiction: the measured, gauge-independent, width 
$\widetilde{\Gamma}$ would differ from the theoretical value
$\Gamma$ by a gauge-dependent quantity in NLO. This contradicts the
premise of the conventional formalism that $\Gamma$, defined in
Eq.~(\ref{eq:usualwidth}), is the radiatively corrected width and is,
furthermore, gauge-independent. We can anticipate that the root of the
problem is that Eq.~(\ref{eq:usualwidth}) is only an approximate
expression for the width of the unstable particle. In particular, it
is not sufficiently accurate when non-analytic contributions are
considered.

It is therefore important to base the calculations in a formalism that
avoids awkward resummations of non-convergent series and the pitfalls
we have encountered in the previous argument.
To achieve this, we return to the transverse dressed $W$ propagator,
inversely proportional to $p^2-M_0^2 -A(p^2)$. In the conventional
mass renormalization one eliminates $M^2_0$ by means of the expression
$M^2_0=M^2 -\mbox{Re}A(M^2)$ (Cf.Eq.~(\ref{eq:M})). 
An alternative possibility is to
eliminate $M^2_0$ by $M^2_0=\sov -A(\sov)$ (Cf.Eq.~(\ref{eq:sov})).
The dressed propagator in the loop integral is inversely proportional
to $p^2 -\sov -[A(p^2)-A(\sov)]$. Expansion of the dressed propagator
leads in Fig.1 to a series in powers of $[A(p^2)-A(\sov)]/(p^2 -\sov)$.
As $A(p^2)-A(\sov)= O[g^2(p^2-\sov)]$ when the loop momentum is in the
resonance region, $[A(p^2)-A(\sov)]/(p^2 -\sov)$ is $O(g^2)$ 
throughout the domain of
integration. Thus, each successive self-energy insertion leads now to
terms of higher order in $g^2$ without awkward non convergent
contributions. In this modified strategy, the zeroth order propagator
in Eq.~(\ref{eq:Wprop-landau}) is replaced by 
\be
        {\cal D}^{\scriptscriptstyle (W,T)}_{\alpha \lambda }(p)=
        \frac{-i}{p^2 -\sov}
        \left(g_{\alpha \lambda} - \frac{p_\alpha p_\lambda}{p^2} \right).
\label{eq:Wprop-landau-sov}
\ee
We note that the imaginary part in $(p^2-\sov)^{-1}$ has the same sign
as Feynman's $i\epsilon$ prescription. Therefore, although the poles of
Eq.~(\ref{eq:Wprop-landau}) in the $k^0$ complex plane are displaced
by the $im_2\Gamma_2$ insertion, they remain in the same quadrants so
that Feynman's contour integration or Wick's rotation can be carried out.
$A_{\smallw \gamma}^{(0)}(s)$, Fig.1 without loop insertions, now
leads directly to 
\be
        A_{\smallw \gamma}^{(0)}(s) = \frac{\alpha}{2\pi}
        \left[(\xig-3)(s-\sov)\LS + \ldots\right].	
\label{eq:A-0-insertions-partial-new}
\ee
$A_{\smallw \gamma}^{(l)}(s)$ ($l\geq 1$), the terms with $l$
insertions in Fig.1, give now contributions of $O(\alpha g^{2l})$, the
normal situation in perturbative expansions. 
The $W$ propagator in the modified formalism is inversely
proportional to $s-\sov- [A(s)-A(\sov)]$. The contribution of 
Eq.~(\ref{eq:A-0-insertions-partial-new}) 
to $[A(s)-A(\sov)]$ is
proportional to $s-\sov$ so that 
the pole position is not displaced, the gauge-dependent contributions
factorize as desired, 
and the previously discussed pitfalls are avoided.
As $A_{\smallw \gamma}^{(l)}(\sov)$ is infrared convergent in the
modified approach, $A_{\smallw \gamma}^{(l)}(s)$ leads now to
contributions to $[A(s)-A(\sov)]$ of order  
$O[(s-\sov)\alpha g^{2l}] = O[\alpha g^{2(l+1)}]$ and can therefore be
neglected in NLO when $l\geq 1$.

The remaining contributions to $A(s)$ from the photonic diagrams,
including those from the longitudinal part of the $W$ propagator in the
loop of Fig.1, and from the diagrams involving the unphysical scalar
$\phi$ and the ghost $C_\gamma$, have no singularities at $s=M^2$
and can therefore be studied with conventional methods. 
In particular, in the evaluation of $A(s)-A(\sov)$ in NLO it is
sufficient to retain their one-loop contributions. In these diagrams
the propagators are proportional to
$(p^2-M^2\xiw)^{-1}$ instead of $(p^2-M^2)^{-1}$. As a consequence,
they lead to logarithmic contributions proportional to 
$$
        (s-M^2)\left[ \frac{s-M^2\xiw}{M^2} \right] \LMxiw,
$$ 
rather than Eq.~(\ref{eq:A-0-insertions-partial}). They have branch
cuts starting at $s=M^2\xiw$, which indicates the unphysical nature of
these singularities. Although they must cancel in physical amplitudes,
they are present in partial amplitudes such as conventional
self-energies and propagators. We briefly discuss how to treat them in
NLO. For $|\xiw-1| \gsim \Gamma/M$, the logarithm can be expanded
about $s=M^2$ and one obtains 
$$
        (s-M^2)\left[1-\xiw + \frac{s-M^2}{M^2} \right] 
        \left[\ln \!\left( \frac{\xiw-1}{\xiw} \right)+
        O \left( \frac{s-M^2}{M^2(1-\xiw)} \right) \right].
$$
The contribution from $O[(s-M^2)/M^2(1-\xiw)]$ is proportional to 
$(s-M^2)^2/M^2$ and is therefore neglected in NLO. For the same
reason, we can neglect $(s-M^2)/M^2$ in the second factor. Therefore,
for $|\xiw-1| \gsim \Gamma/M$, in NLO we can approximate this
contribution by the simple expression
$(s-M^2)(1-\xiw)\ln[(\xiw-1)/\xiw]$.
For $|\xiw-1| \lsim \Gamma/M$ the expansion of the logarithmic factor
is not valid, but we note that the whole contribution is proportional
to $(s-M^2)^2$ or $(s-M^2)(1-\xiw)$ and therefore negligible in
NLO. As a consequence, the above mentioned approximation can be used
for any value of $\xiw$. 
Calling $A^{\gamma}(s)$ the overall contribution of the 
one-loop photonic diagrams to the transverse $W$ self-energy (Fig.2),
in the modified
formulation the relevant quantity in the correction to the $W$
propagator is $A^{\gamma}(s) - A^{\gamma}(\sov)$. In general 
$R_\xi$ gauge, we find in NLO
\bea 
        \lefteqn{\Agamma(s) - \Agamma(\sov) = 
                \frac{\alpha(m_2)}{2\pi} (s-\sov)
                \left\{\delta 
                \left(\frac{\xiw}{2}-\frac{23}{6}\right)
                +\frac{34}{9} -2\LS
                                \right.}                \nonumber \\
        & & \!\!\!\!\!\!\!\!\!\!
        -\left(\xiw-1\right)\left[\fr{\xiw}{12} -
                \left(1-\fr{(\xiw-1)^2}{12}\right)
                \ln \!\left(\fr{\xiw-1}{\xiw}\right) \right]      
                -\left(\fr{11}{12} 
                -\fr{\xiw}{4} \right) \ln\xiw
							\nonumber \\
        & & \!\!\!\!\!\!\!\!\!\!
        \left. +\left(\xig-1\right)\left[
                \fr{\delta}{2}+\fr{1}{2}
                +\LS+\fr{(\xiw^2-1)}{4}\ln\!\left(\fr{\xiw-1}{\xiw}\right) 
                -\fr{\ln\xiw}{4}+\fr{\xiw}{4} \right] \right\},
\label{eq:Agamma-Agamma}                                                       
\eea
where $\delta = (n-4)^{-1} + (\gamma_E -\ln 4\pi)/2$, we have
treated the logarithmic terms according to the previous discussion
and set $\mu=m_2$. The corresponding one-loop gluonic contribution to
the quark self-energy is depicted in Fig.3.

Writing 
\be
        1-\fr{s}{\sov} = 1-\fr{s}{m_1^2}-i\fr{s}{m_1^2}
        \fr{\Gamma_2}{m_2} = \rho e^{i\theta},
\ee
we have
\be
        \rho = \left[\left(1-\fr{s}{m_1^2}\right)^2 +
        \fr{s^2 \Gamma_2^2}{m_1^4 m_2^2} \right]^{1/2}, 
\label{eq:rho}
\ee
\be
        \rho \sin \theta = -\fr{s \Gamma_2}{m_1^2 m_2},
\label{eq:rhosintheta}
\ee
where $m_1$ is defined in Eq.~(\ref{eq:m1}).
Calling $\alpha \equiv \sin^{-1}(\Gamma_2/m_1)$, we have:
for $-\infty <s<0$, $\alpha>\theta>0$;
for $0<s<m_1^2$, $0>\theta>-\pi/2$;
for $m_1^2<s<\infty$, $-\pi/2>\theta>-\pi+\alpha$.
In Figs.4 and 5 the functions $\ln\rho(s)$ and $\theta(s)$ are plotted 
for $m_1=80.4\gev$ and $\Gamma_1=\Gamma_2 m_1/m_2=2\gev$
over a large range of $\sqrt{s}$ values. 
Figs.6 and 7 compare these functions 
with the zero-width approximations over the resonance region. 
In the limit $\Gamma_2 \rightarrow 0$, 
$\theta(s)$ becomes a step function. This corresponds to the
well-known expression 
$$
        \mbox{Im}\left[\ln \!\left(\fr{M^2-s -i\epsilon}{M^2} 
        \right)\right]=  - \pi \theta\left(s-M^2\right),
$$
where the $i\epsilon$ prescription implies $\theta(0)=1/2$.
The zero width approximation, however, is not valid in the resonance
region.

Eq.~(\ref{eq:Agamma-Agamma}) exhibits a number of interesting
theoretical features: a) the coefficient of $\ln[(\sov-s)/\sov]$ is
independent of $\xiw$ but is proportional to $(\xig-3)$. b) The
logarithm $\ln(\xiw-1)$ contains an imaginary contribution 
$-i\pi \theta(1-\xiw)$. This can be understood from the observation
that, for $\xiw<1$, a $W$ boson of mass $s=M^2$ has non-vanishing
phase space to ``decay'' into a photon and particles of mass $M^2\xiw$.
As explained before, Eq.~(\ref{eq:Agamma-Agamma}) is only valid in the 
resonance region.

For completeness, the full one-loop expression for $\Agamma(s)$ in
general $R_\xi$ gauges is given in the Appendix.

\section{Fried-Yennie Gauge and the PT Prescription. 
\\Gauge Dependence of the On-Shell Mass}

We note that the $\ln [(\sov-s)/\sov]$ terms in 
Eq.~(\ref{eq:A-0-insertions-partial-new})
and Eq.~(\ref{eq:Agamma-Agamma}) cancel for $\xig=3$, the gauge
introduced by Fried and Yennie  in Lamb-shift 
calculations \cite{Fried-Yennie}. 
It should be emphasized, however, that a gauge-independent logarithm 
of this type survives in
physical amplitudes involving unstable particles such as
$W$ \cite{HoWa}.
Thus, the choice $\xig=3$ removes this contribution from the 
propagator's corrections, but not the overall amplitude. In this
connection, it is interesting to inquire how the Pinch Technique (PT)
prescription treats these terms. We recall that the PT is a
prescription that combines the conventional self-energies with ``pinch
parts'' from vertex and box diagrams in such a manner that the
modified self-energies are independent of $\xi_i$ $(i=W, \gamma, Z)$
and exhibit desirable theoretical properties. Calling $a(q^2)$ the PT
$W$ self-energy, we recall that in the Standard Model (SM)
\be
        a(s) = \left[ A(s) \right]_{\xi_i=1} -
        4g^2(\mu) (s-M^2)\left[\cos^2 \theta_\smallw I_{\smallw \smallz}(s) +
        \sin^2 \theta_\smallw I_{\gamma \smallw}(s) \right],
\ee
where
$$
        I_{ij}(s)= i \mu^{4-n} \int \frac{d^n \!k}{(2\pi)^n}
        \fr{1}{(k^2-m_i^2)[(k+q)^2-m_j^2]},
$$
and tadpole contributions have been included in both $a(s)$ and
$A(s)$ \cite{DeSi92}.
The $I_{\gamma \smallw}(s)$ term leads to a contribution 
$-(\alpha/\pi)[(s-M^2)^2/s]\ln[(M^2-s)/M^2]$, which is of higher order
in $(s-M^2)$. Therefore, in NLO the PT self-energy generates the same 
$\ln [(\sov-s)/\sov]$ term as the 't Hooft--Feynman gauge $(\xig=1)$,
i.e. $-(\alpha/\pi)(s-\sov) \ln [(\sov-s)/\sov]$. The possibility has
been suggested in the past to define the on-shell mass in terms of the
PT self-energy, namely $M^2 =M_0^2 + \mbox{Re} \, a(M^2)$ \cite{PhSi96}.
This has the
advantage that one is dealing here with a $\xi_i$-independent
amplitude. Repeating the argument after 
Eq.~(\ref{eq:A-0-insertions-partial-again}), we see however that
conventional on-shell renormalization based on $a(s)$ would lead to a
contribution $iM\Gamma(\alpha/\pi)[1+i\pi/2]$ which,
although $\xi_i-$independent, is not
proportional to the zeroth order term $s-M^2 +iM\Gamma$. Its
removal would require a redefinition of $M$ and $\Gamma$, which is
inconsistent with the fact that $\Gamma$ contains all the corrections
of $O(\alpha)$. 
This problem can be circumvented once more by
recalling that the PT does not displace the position of the complex
pole at least through $O(g^4)$ \cite{PaPi-PhSi}, 
and expressing the inverse propagator
as $s-\sov - [a(s)-a(\sov)]$. The contribution of the 
$(s-\sov) \ln [(\sov-s)/\sov]$ terms to $a(s)-a(\sov)$ is proportional
to $(s-\sov)$ and the above mentioned difficulties are avoided.

The difference between $m_1$, defined in Eq.~(\ref{eq:m1}), and the
conventional on-shell mass $M$, defined in  Eq.~(\ref{eq:M}), is
\be
        M^2 -m_1^2 = \mbox{Re} A(M^2) - \mbox{Re} A(\sov) -\Gamma_2^2.
\label{eq:deltaM}
\ee
The contribution of the $(s-\sov)\ln[(\sov-s)/\sov]$ term to the r.h.s. 
of Eq.~(\ref{eq:deltaM}) is
\bea
	& &\frac{\alpha(m_2)}{2\pi}\left(\xig -3\right)
	   \left[\left(M^2-m_2^2\right) 
		\mbox{Re}\ln \!\left(\frac{\sov-M^2}{\sov}\right)
		-m_2 \Gamma_2 \mbox{Im}\ln \! 
		\left(\frac{\sov-M^2}{\sov}\right)
	   \right]				\nonumber \\
	& \approx & \frac{\alpha(m_2)}{2\pi}\left(\xig -3 \right)
	   \left[\left(M^2-m_1^2\right)
		\mbox{Re}\ln \! \left(\frac{\sov-M^2}{\sov}\right)
		+m_2 \Gamma_2 \frac{\pi}{2}\right].
\eea
In $\mbox{Im}\ln[(\sov-M^2)/\sov]$ we have approximated 
$M^2 \approx m_1^2$ and used the fact that $\theta=-\pi/2$ for
$s= m_1^2$ (see discussion after Eq.~(\ref{eq:rhosintheta})).
Thus, we see that 
\be
        M^2 -m_1^2 = \fr{\alpha(m_2)}{4}(\xig-3)m_2\Gamma_2+ \ldots,
\label{eq:deltaMagain}
\ee
where the dots indicate additional contributions. Note that this last
equation corresponds to Eq.~(\ref{eq:Mtilde}) with the identification
$\widetilde{M} \rightarrow m_1$.

As $\xig$ can be arbitrarily large, Eq.~(\ref{eq:deltaMagain}) reveals that
in the $W$ case the gauge dependence of the conventional on-shell
definition of mass is unbounded in NLO for any value of $\xiw$. 
Similarly, the term proportional to 
$(s-\sov)(\xig-1)(\xiw^2-1)\ln(\xiw-1)$ in
Eq.~(\ref{eq:Agamma-Agamma}) gives an unbounded contribution 
$(\alpha/8)(\xig-1)M\Gamma(\xiw^2-1)\,\theta(1-\xiw)$ to 
$M^2-m_1^2$ in the restricted range $\xiw<1$.
This situation is to be contrasted with
the $Z$ case, where the gauge dependence in NLO is bounded and 
$\lsim 2\mev$ \cite{PaSi96}.
The difference is due to the contribution of the
logarithms from the photonic diagrams, which are absent in the $Z$
case. In particular, in the frequently employed 't Hooft--Feynman
gauge $(\xi_i=1)$, Eq.~(\ref{eq:deltaMagain}) leads to 
$m_1-M=\alpha(m_2) \Gamma_2/4 \approx 4$ MeV. In analogy with the $Z$ case, 
there are also bounded gauge-dependent contributions to $m_1-M$
arising from non-photonic diagrams in the restricted range
$\sqrt \xiz \leq \cos \theta_{\mbox{\footnotesize{w}}}[1-\sqrt \xiw]$
and from the photonic corrections proportional to 
$(\xiw-1)\ln[(\xiw-1)/\xiw]$ (Cf. Eq.~(\ref{eq:Agamma-Agamma})).

\section{Overall Corrections to \boldmath{$W$} Propagators in the 
Resonance Region}

In contrast with the photonic corrections, the non-photonic
contributions $A_{np}(s)$ to $A(s)$ are analytic around $s=\sov$.   
We can therefore write 
\be
	A_{np}(s)-A_{np}(\sov) = (s-\sov)A_{np}'(m_2^2)+ \ldots,
\ee
where the dots indicate higher-order contributions. 

In the resonance region, and in NLO, the transverse $W$ propagator 
is given by 
\be
	{\cal D}^{\scriptscriptstyle (W,T)}_{\alpha \beta}(q)=
        \frac{-i \left(g_{\alpha \beta} - q_\alpha q_\beta/ q^2\right)}
	     {\left(s-\sov\right)
		\left[1-A_{np}'(m_2^2) 
		-\frac{\alpha(m_2)}{2\pi}F(s,\sov,\xig,\xiw) \right]},
\label{eq:fullWprop-landau}
\ee
where $s=q^2$ and $F(s,\sov,\xig,\xiw)$ is the expression between
curly brackets in Eq.~(\ref{eq:Agamma-Agamma}). An alternative
expression, involving an $s-$dependent width, can be obtained by
splitting $A_{np}'$ into real and imaginary parts, and the latter into 
fermionic Im$A'_f$ and bosonic Im$A'_b$ contributions. 
Neglecting very small scaling violations, we have 
\be
	\mbox{Im} A'_f(m_2^2) \approx \mbox{Im} A_f(m_2^2)/m_2^2
			\approx -\Gamma_2/m_2.
\ee
Eq.~(\ref{eq:fullWprop-landau}) becomes then 
\be
	{\cal D}^{\scriptscriptstyle (W,T)}_{\alpha \beta}(q)=
        \frac{-i \left(g_{\alpha \beta} - q_\alpha q_\beta/ q^2\right)}
	     	{\left(s-m_1^2 +is\frac{\Gamma_1}{m_1}\right)
	     	\left[1-\mbox{Re}A_{np}'(m_1^2)-i\mbox{Im}A_b'(m_1^2)
		-\frac{\alpha(m_1)}{2\pi}F \right] },
\label{eq:fullWprop-landau-again}
\ee
where $\Gamma_1/m_1=\Gamma_2/m_2$. $\mbox{Im}A_b'(m_1^2)$ is non zero
and gauge-dependent in the subclass of gauges that satisfy 
$\sqrt \xiz \leq \cos \theta_{\mbox{\footnotesize{w}}}[1-\sqrt \xiw]$.
(If this condition is satisfied, a $W$ boson of mass $\sqrt s \approx \mw$ 
has non-vanishing phase space to ``decay'' into particles of mass
$\mw \sqrt \xiw$ and $\mz \sqrt \xiz$.) Otherwise $\mbox{Im}A_b'(m_1^2)$
vanishes. Although $m_1$ and $\Gamma_1$ are gauge-invariant, 
$\mbox{Re}A'_{np}(m_1^2)$, $\mbox{Im}A'_{np}(m_1^2)$
 and $F$ are gauge-dependent. In physical
amplitudes, such gauge-dependent terms cancel against contributions
from vertex and box diagrams.
The crucial point is that the gauge-dependent contributions in 
Eq.~(\ref{eq:fullWprop-landau-again}) factorize so that such
cancelations can take place and the position of the complex pole is
not displaced.

\section{Comparison of the \boldmath{$W$} Width in the Conventional
and Modified Formulations}

In this Section we show that the conventional and modified
formulations lead to the same result for the $W$ width in NLO.
However, the two approaches differ in higher orders.
In particular, the conventional formulation is plagued in high
orders by severe infrared singularities. 
Calling $A_0(s, M^2_0)$ the transverse self-energy evaluated in terms
of the bare mass $M_0$, and $A(s,M^2)$ and $\Aov(s,\sov)$ the
expressions obtained by substituting
$M_0^2=M^2-\mbox{Re}A(M^2,M^2)$ and $M_0^2=\sov -\Aov(\sov,\sov)$,
respectively, we have 
\be
	A_0(s, M^2_0) = A(s,M^2) = \Aov(s,\sov).
\label{eq:AAA}
\ee
In the conventional approach the $W$ width is given by  
Eq.~(\ref{eq:usualwidth}) or, equivalently, 
\be
	M\Gamma = -\mbox{Im} A(M^2,M^2) + M \Gamma \,\mbox{Re}A'(M^2,M^2),
\label{eq:usualwidth-again}
\ee
where the prime means differentiation with respect to the first argument.
Instead, in the modified formulation discussed in the present paper,
the width is defined by 
\be
	m_2 \Gamma_2 = -\mbox{Im} \Aov (\sov,\sov),
\label{eq:newwidth}
\ee
which follows from Eq.~(\ref{eq:sov}). Combining Eq.~(\ref{eq:newwidth})
with Eq.~(\ref{eq:AAA}) we find:
\bea
	m_2 \Gamma_2 &=& -\mbox{Im} A (\sov,M^2) \nonumber  \\
		   &=& -\mbox{Im} A (M^2,M^2)
		       -\mbox{Im} \left[\left(
			\sov-M^2\right) A'(M^2,M^2) \right] + O(g^6).
\label{eq:newwidth-again}
\eea
As $\sov -M^2 = m_2^2-M^2 -im_2\Gamma_2$ and $m_2^2-M^2=O(g^4)$, 
Eq.~(\ref{eq:newwidth-again}) becomes 
\be 
	m_2 \Gamma_2 = -\mbox{Im} A (M^2,M^2) +
		m_2\Gamma_2 \,\mbox{Re}A'(M^2,M^2)+ O(g^6).
\label{eq:newwidth-again2}
\ee
Comparing Eq.~(\ref{eq:usualwidth-again}) and Eq.~(\ref{eq:newwidth-again2}) 
we see that indeed 
\be
	\Gamma_2 = \Gamma + O(g^6).
\ee
Thus, the two calculations of the width coincide through $O(g^4)$,
i.e. in NLO. It is interesting to see how the two formulations treat
potential infrared divergences. As is well-known, 
$\mbox{Re} A_{\gamma}'(M^2,M^2)$, the photonic contribution to 
$\mbox{Re} A'(M^2,M^2)$, is logarithmically infrared
divergent. Therefore, $M \Gamma \mbox{Re}A'(M^2,M^2)$ in the
last term of Eq.~(\ref{eq:usualwidth-again}) contains a logarithmic
infrared divergence in $O(\alpha g^2)$. This is canceled by an
infrared divergence in $\mbox{Im} A(M^2,M^2)$ arising from 
$A_{\smallw \gamma}^{(1)}(M^2,M^2)$, i.e. Fig.1 with one self-energy
insertion. As it is clear from the discussion of Section 2, the
infrared divergence in $A_{\smallw \gamma}^{(1)}(M^2,M^2)$ has its
origin in the fact that the self-energy insertion induces a correction
factor $i\mbox{Im} A(M^2)/(p^2-M^2)$, where $p$ is the $W$ loop
momentum. 

In higher orders the problem of infrared divergences in the
conventional approach becomes severe. It follows from 
Eq.~(\ref{eq:A-l-insertions-Im-again}) that the diagrams in Fig.1
generate infrared divergences of 
$O[\alpha (-i)^l M \Gamma (\Gamma/\lambda_{min})^{l-1}]$ in
$A(M^2,M^2)$, where $\lambda_{min}$ is the infrared cut-off.
As a consequence, Eq.~(\ref{eq:usualwidth-again}), the width evaluated
in the conventional formulation, contains a power-like infrared divergence
of $O[\alpha (\xig-3) M\Gamma (\Gamma/\lambda_{min})^2]$ which appears
in $O(\alpha g^6)$. Similarly, the conventional mass renormalization
counterterm $\delta M^2= \mbox{Re}A(M^2,M^2)$ contains an infrared 
divergence of $O[\alpha (\xig-3) M\Gamma^2/\lambda_{min}]$ that appears
in $O(\alpha g^4)$. One can avoid these leading infrared divergences
by resumming the contributions of the 
$\mbox{Im}A(M^2,M^2) \approx -M \Gamma$ insertions in Fig.1. As
explained in Section 2, this would lead to the replacement
$$
	\frac{\alpha}{2\pi}(\xig-3)(s-M^2)\LM 
	\rightarrow
	\frac{\alpha}{2\pi}(\xig-3)(s-M^2+iM\Gamma)\
		\ln \! \left(\frac{M^2-s-iM\Gamma}{M^2-iM\Gamma} \right).
$$
Unfortunately, the contribution of this resummed expression to the
r.h.s. of Eq.~(\ref{eq:usualwidth-again}) is
$(\alpha/2\pi)(\xig-3)M\Gamma$, a gauge-dependent contribution of 
$O(\alpha g^2)$ to the width! In contrast, in the modified formulation
the corresponding expression is 
$(\alpha/2\pi)(\xig-3)(s-\sov)\ln[(\sov-s)/\sov]$ and causes no
problem since it gives no contribution to Eq.~(\ref{eq:newwidth}).
It is also important to note that $\Aov_{\smallw \gamma}(\sov,\sov)$
is infrared convergent in all orders, since the self-energy insertions
induce a correction factor 
$[\Aov(p^2,\sov)-\Aov(\sov,\sov)]^l/(p^2-\sov)^l$ in the integrand of
Fig.1, and this converges, modulo logarithms, as $p^2\rightarrow \sov$.
In particular, it is easy to check that the contributions of
$O[\alpha (\xig-3) M\Gamma (\Gamma/\lambda_{min})^2]$ to the width
mentioned above are canceled by terms of $O(\alpha g^6)$ in the
expansion of Eq.~(\ref{eq:newwidth-again}). In the conventional
formulation such terms are not included 
(Cf. Eq.~(\ref{eq:usualwidth-again})) and this leads to the problem of
uncompensated infrared singularities  in high orders of perturbation
theory. Other theoretical difficulties of the conventional definition 
of width and the need to replace it by 
Eq.~(\ref{eq:newwidth})
have been emphasized in 
Ref.~\cite{BhWi} and Ref.~\cite{Si-Ring}.

\section{QCD Corrections to Quark Propagators in the Resonance Region}

In pure QCD quarks are stable particles. However, they become unstable
when weak interactions are switched on. An example of a reaction that
may probe the top quark propagator in the resonance region is 
$W^+ + b \rightarrow t \rightarrow W^+ + b$. In this Section we discuss
in NLO the QCD part of the corrections to the quark propagator in the
resonance region. The relevant diagram is depicted in Fig.3. Because
the gluons are massless, we anticipate problems analogous to those
discussed in Section 2. Therefore, we work from the outset in the
complex pole formulation. Denoting the position of the complex pole by
$\mov = m -i \Gamma/2$, we observe that $\Gamma$ arises from the weak 
interactions. For example, in the top case $\Gamma$ emerges in lowest
order from the imaginary part of the $W b$ and $\phi b$ contributions
to the top self-energy. If we treat $\Gamma$ in lowest order in the weak 
interactions, but otherwise neglect the remaining weak interaction 
contributions to the self-energy, the dressed quark propagator can be
written as 
\be
	S'_F(\qsl) = \frac{i}{\qsl -\mov - 
		\left( \Sigma(\qsl) - \Sigma(\mov) \right)},  
\ee
where $\Sigma(\qsl)$ is the pure QCD contribution.

Decomposing 
\be
	\Sigma(\qsl) = \mov A(q^2) + \qsl B(q^2),
\ee 
and using $i/(\psl-\mov)$ as loop propagator, we find from Fig.3:
\bea
	A(q^2) &=& \frac{\alpha_s(m)}{3\pi}
		\left\{ -2+ \left(\xi_g+3 \right) \left[
		-2 \delta +2 + \left(\frac{\mov^2 -q^2}{q^2}\right)
		\ln \! \left(\frac{\mov^2 -q^2}{\mov^2}\right)
		 \right] \right\}, 
						 \\
	B(q^2) &=& \xi_g \frac{\alpha_s(m)}{3\pi}
		\left\{ 2 \delta -1 - \frac{\mov^2}{q^2} -
		\left(\frac{\mov^4 -q^4}{q^4}\right)
		\ln \! \left(\frac{\mov^2 -q^2}{\mov^2}\right)
		\right\}, 
\label{eq:QCD-self-en}
\eea
where $\xi_g$ is the gluon gauge parameter and we have set $\mu=m$.
In NLO in the resonance region this simplifies to 
\be
	\Sigma(\qsl) = \frac{\alpha_s(m)}{3\pi} \left\{(\qsl -\mov) 
		\left[2\left(\xi_g-3\right) 
			\ln \! \left(\frac{\mov^2 -q^2}{\mov^2}\right)
			+2 \delta \xi_g \right]+ 
		\mov \left[ 4-6\delta \right] \right\} + \ldots
\label{eq:QCD-self-en-approx}
\ee
and
\be
	S'_F(\qsl) = \frac{i}{\left(\qsl -\mov \right)} 
		\left\{1-
		\frac{\alpha_s(m)}{3\pi} \left[2\left(\xi_g-3\right) 
			\ln \! \left(\frac{\mov^2-q^2}{\mov^2}\right)+
			2 \delta \xi_g \right]+\ldots \right\}^{-1}.
\label{eq:S'_F}
\ee
As in the $W-$propagator case, we see that the logarithm vanishes in
the Fried--Yennie gauge $\xi_g =3$. In fact, its coefficient can be
obtained from the analogous term in 
Eq.~(\ref{eq:A-0-insertions-partial-new})
by substituting $\alpha \rightarrow (4/3)\alpha_s(m)$, where $4/3$
arises from the color factor. Writing once more $s=q^2$, 
$1-s/\mov^2=\rho(s)e^{i\theta(s)}$, the functions $\rho(s)$ and 
$\theta(s)$ are given by Eq.~(\ref{eq:rho}) and 
Eq.~(\ref{eq:rhosintheta}) with the identification
$m_2 = (m^2 -\Gamma^2/4)^{1/2}$, 
$\Gamma_2 =m\Gamma/m_2$, and 
$m_1$ defined in Eq.~(\ref{eq:m1}). 
The difference between $m$ and the on-shell mass
$M=m_0 +\mbox{Re}\Sigma(M)$ in leading order is 
\be 	
	M-m = -\frac{\alpha_s(m)}{3\pi} \Gamma \left(\xi_g-3\right)
		\mbox{Im} \ln \! \left(\frac{\mov^2-M^2}{\mov^2}\right)
	    = \frac{\alpha_s(m)}{6}\Gamma \left(\xi_g-3\right),
\label{eq:deltaM-QCD}
\ee
which can also be obtained from Eq.~(\ref{eq:deltaMagain}) by
substituting once more $\alpha(m_2) \rightarrow (4/3)\alpha_s(m)$.
Thus, in analogy with the $W$ case, $M-m$ is unbounded in NLO. In the
Feynman gauge $(\xi_g=1)$ Eq.~(\ref{eq:deltaM-QCD}) leads to 
$m-M=\alpha_s(m)\Gamma/3 \approx 56\mev$, while in the Landau gauge
($\xi_g=0$) we have $m-M \approx 84\mev$.

\section{Conclusions}

We have shown in Section 2 that conventional mass renormalization 
(Eq.~(\ref{eq:M})), when applied to the photonic and gluonic diagrams,
leads to a series in powers of $M\Gamma/(s-M^2)$, which does not
converge in the resonance region
(Eq.~(\ref{eq:A-l-insertions-Im-again})). 
In Section 5 we have pointed out that this behavior induces in high
orders power-like infrared divergences in both $M$ and $\Gamma$.
In principle, these severe problems can be circumvented by a
resummation procedure, explained in Section 2. Unfortunately, the
resummed expressions are incompatible with the conventional definition
of width (Eq.~(\ref{eq:usualwidth}) and
Eq.~(\ref{eq:usualwidth-again})).  
In fact, combining the resummed expression with these equations, we
have encountered gauge-dependent corrections of $O(\alpha \Gamma)$ to
the width and resonant propagator, in contradiction with basic
theoretical properties of these amplitudes.  This clash between the
resummed expressions and the conventional definition of width is not
difficult to understand. Indeed, the usual derivation of the latter
treats the unstable particle as an asymptotic state, which is clearly
an approximation. In Section 2 and 5 we have discussed an alternative
treatment of the resonant propagator and the width based on the
complex pole position $\sov=M_0^2 +A(\sov)$. The non-convergent terms
in the resonant region and the potential infrared divergences in
$\Gamma$ and $M$ are avoided by employing $(p^2-\sov)^{-1}$ rather than
$(p^2 -M^2)^{-1}$ in the Feynman integrals, where $p$ is the $W$ or
quark loop momentum. The one-loop diagrams 
lead now directly to the 
resummed expression of the conventional approach, while the multi-loop
expansion generates terms which are genuinely of higher order.
The non-analytic terms and gauge-dependent corrections in the resonant
region cause no problem because they are proportional to $s-\sov$ and
exactly factorize. 
We emphasize that this is a crucial property, since it implies that
the pole position is not displaced and the gauge-dependent corrections
can be canceled by vertex and box contributions.
Furthermore, they do not lead to difficulties in
the evaluation of the width because the latter is now defined by 
Eq.~(\ref{eq:newwidth}). In particular, the non-analytic contributions
cancel exactly in its evaluation and the answer is infrared convergent
to all orders in the perturbative expansion. Comparing the masses
defined in the two approaches, in Section 3 we have reached the
conclusion that, unlike the $Z$ case, the gauge dependence of the
conventional definition of mass (Eq.~(\ref{eq:M})) is unbounded in NLO
for any value of $\xiw$. In Section 5 it is further shown that the
conventional and alternative formulations of the width coincide in
NLO, but not beyond. The analysis reveals also a curious and perhaps
universal property: in NLO the non-analytic terms in both the $W$ and
quark propagators vanish in the Fried--Yennie gauge $\xig=3$.

In the past, a number of authors have employed heuristically the
replacement 
$\ln[(M^2-s)/M^2]\rightarrow \ln[(M^2-iM\Gamma-s)/(M^2-iM\Gamma)]$ in
order to avoid apparent on-shell singularities
(see, for example, Refs.~\cite{HoWa,BDP} and the first article
of Ref.~\cite{var}).
In this paper we have
attempted to clarify the theoretical basis for this heuristic
procedure and shown how it emerges from the formalism. In fact, the 
analysis leads to the conclusion that the replacement 
$M^2 \rightarrow \sov$ must be made in the complete expression of the
non-analytic terms and that, at the same time, the definition of width
must be changed from Eq.~(\ref{eq:usualwidth-again}) to 
Eq.~(\ref{eq:newwidth}).

The idea of employing $\sov$, rather than the conventional approach,
as a basis to define the mass and width of unstable particles and
analyze the propagator in the resonance region has been recently
advocated, for different theoretical reasons, by a number of
theorists \cite{Si91a,Si91b,PaSi96,BhWi,Si-Ring,var}. 
The arguments in this paper provide an
additional and powerful argument for such approach.


\vspace{1cm}
\begin{center}
{\large\bf Acknowledgments}
\end{center}
\bigskip
\noindent
We would like to thank B.~A.~Kniehl, M.~Porrati, and M.~Schaden
for very useful discussions. This research was supported in part by
NSF Grant PHY-9722083.

\newpage
\setcounter{section}{1}	\setcounter{equation}{0}
\def\theequation{\Alph{section}.\arabic{equation}}
\section*{Appendix: Photonic Corrections to the Transverse
\boldmath{$W$} Self-Energy in General \boldmath{$R_{\xi}$} Gauges}

The conventional evaluation of the contribution of Fig.2 to
the transverse $W$ self-energy is given by
\be
        \Agamma(s) = \Agamma(s)|_{\xi_i=1} + 
		\Delta \Agamma(s)|_{\xi_i \neq 1},
\ee
where $\Agamma(s)|_{\xi_i=1}$, the contribution in the 
't Hooft--Feynman gauge, is
\bea
        \Agamma(s)|_{\xi_i=1} &=& \fr{\alpha}{2\pi} 
                \left\{-\left(\delta +\ln \fr{M}{\mu}\right)
		\left(\frac{10}{3}s+3M^2\right) 
                +\frac{11}{6}M^2 +\fr{31}{9}s
                -\fr{M^4}{3s} 
                                        \right. \nonumber \\
        & & \;\;\;\;\;\;\;\,                        \left. 
                +\frac{1}{3}\left(M^2-s\right) L(s,M^2)
                \left[5 +\fr{2M^2}{s} -\fr{M^4}{s^2}
                                \right] \right\}. 
\eea
The remainder is
\bea
        \Delta \Agamma(s)|_{\xi_i \neq 1} &=& \fr{\alpha}{4\pi} \left\{
                \;L(s,M^2) \fr{\left(M^2-s\right)^2}{M^2}
                \left[\fr{M^2+s}{s} - \fr{(M^2-s)^2}{6s^2}\right]
                                                \right. \nonumber  \\
        & &    	\!\!\!\!\!\!\!\!\!\!\!\!\!\!\!\!\!\!\!\!\!\!\!\!\!\!\!\!
                \!\!\!\!\!\!\!\!
		-L(s,M^2\xiw) \left(M^2-s\right) 
                \fr{\left(M^2\xiw-s\right)}{M^2}
                \left[ \fr{M^2+s}{s} - \fr{(M^2\xiw-s)^2}{6s^2}\right]
                                                        \nonumber  \\
        & &    	\!\!\!\!\!\!\!\!\!\!\!\!\!\!\!\!\!\!\!\!\!\!\!\!\!\!\!\!
                \!\!\!\!\!\!\!\!
		+\left(\xig-1\right)\left(\fr{M^2-s}{4}\right) \!\!
                \left[-4\left(\delta +\ln\fr{M}{\mu}\right)
		-\fr{M^2}{s}\left(\xiw+1\right)
                +\left(1+\fr{s}{M^2}\right)\ln \xiw
                                \right.   \nonumber  \\
        & &   	\!\!\!\!\!\!\!\!\!\!\!\!\!\!\!\!\!\!\!\!\!\!\!\!\!\!\!\!
		\!\!\!\!\!\!\!\!\left.
                -\xiw -3 -\left(\fr{M^2+s}{s^2 M^2}\right) \!\!
                \left( \phantom{\fr{M^2}{M^2}}\!\!\!\!\!\!\!\!\!\!
		\left(M^2+s\right)^2 L(s,M^2) +
                \left(M^4 \xiw^2-s^2\right) L(s,M^2\xiw)
                \right) \right]   \nonumber    \\
        & &    	\!\!\!\!\!\!\!\!\!\!\!\!\!\!\!\!\!\!\!\!\!\!\!\!\!\!\!\!
                \!\!\!\!\!\!\!\!
		+\left(\xiw-1\right) \left[
                \left(\delta +\ln\fr{M}{\mu}\right)
		\left(s -\fr{3}{2}M^2\xiw - \fr{M^2}{2} \right)
                +\xiw\left(\fr{11}{24}M^2+\fr{M^4}{6s}\right)+\fr{s}{6}
                	        \right. \nonumber    \\
        & &    	\!\!\!\!\!\!\!\!\!\!\!\!\!\!\!\!\!\!\!\!\!\!\!\!\!\!\!\!
		\!\!\!\!\!\!\!\!\left. \left.
                -\fr{17}{24}M^2 +\fr{M^4}{6s} \right]
                + \ln\xiw \left[M^2 -\fr{s}{6}-\fr{5s^2}{6M^2}
                +\fr{\xiw}{2}
                \left(M^2+s\right) -\fr{3}{4}M^2\xiw^2 \right] \right\},
                                                \nonumber    \\
        & & 
\label{eq:DeltaAgamma}
\eea
where 
$$
        L(x,y) = \ln\left(\fr{y-x}{y}\right),
$$
and $\delta$ is defined after Eq.~(\ref{eq:Agamma-Agamma}).

The $\xi_i-$dependence of the complete transverse self-energy $A(s)$
must vanish on-shell (provided the tadpole contributions are
included). The photonic diagrams give rise to all the $\xig-$dependent
and $L(s,M^2\xiw)$ contributions in $A(s)$, and we see that these
terms indeed cancel when $s=M^2$. The non-vanishing terms in 
Eq.~(\ref{eq:DeltaAgamma}) are canceled on-shell by non-photonic
contributions.


\newpage
\section*{Figures}


$$\epsfig{figure=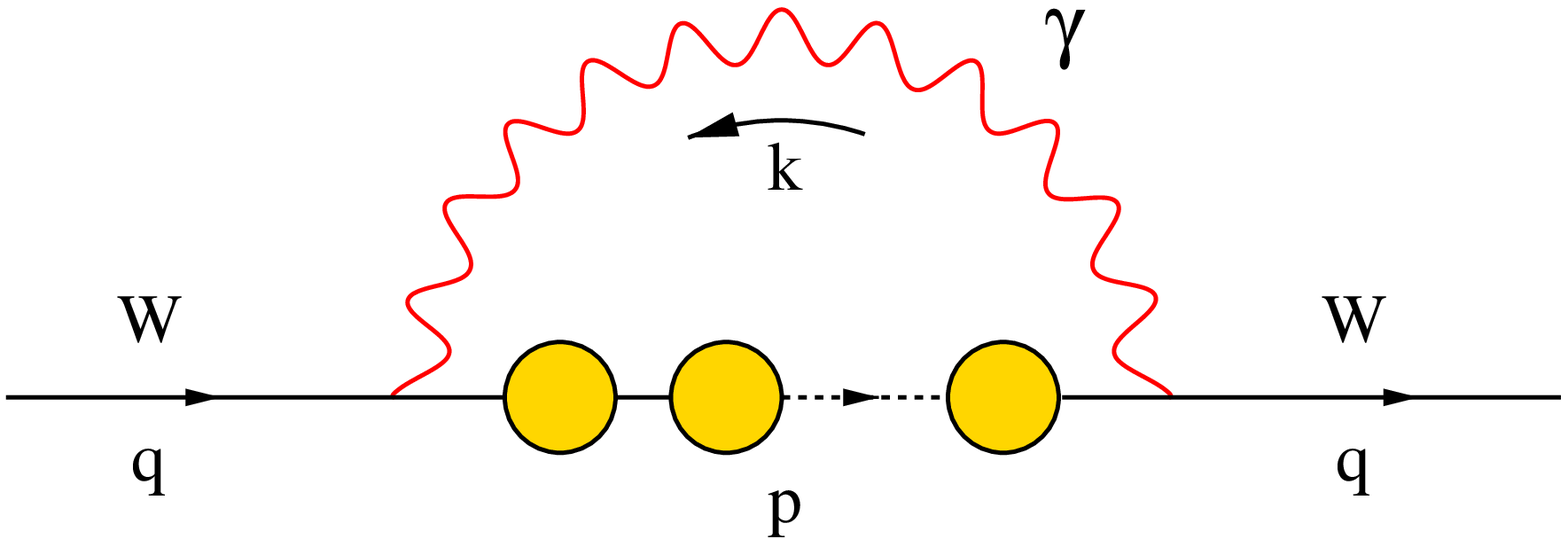,width=15.3cm}$$
Fig.1 A class of photonic corrections to the $W$ self-energy.
The inner solid and dashed lines and blobs represent transverse 
$W$ propagators and self-energies. 

$$\epsfig{figure=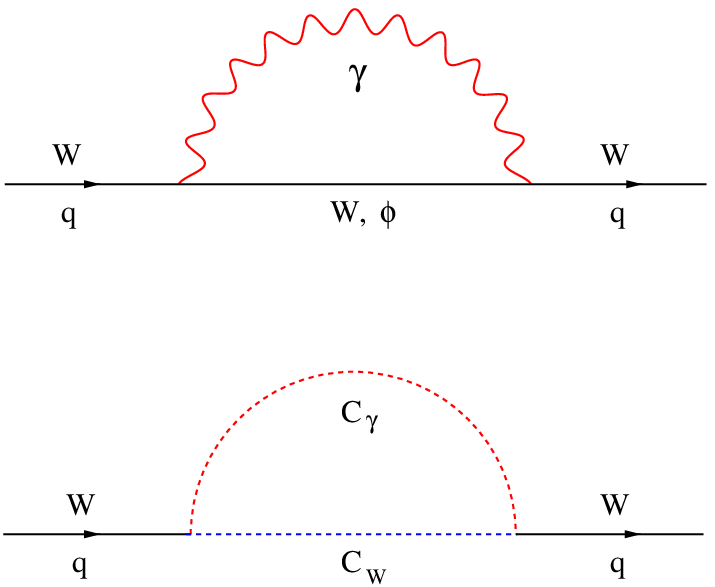,width=8cm}$$
Fig.2 One-loop photonic diagrams for the $W$ 
self-energy; $\phi$ is the unphysical scalar, $C_\gamma$ and 
$C_W$ are ghosts.

$$\epsfig{figure=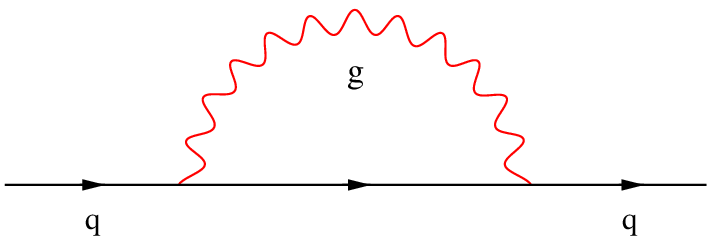,width=8cm}$$
\nobreak
Fig.3 One loop diagram for the quark self-energy in QCD.

$$\epsfig{figure=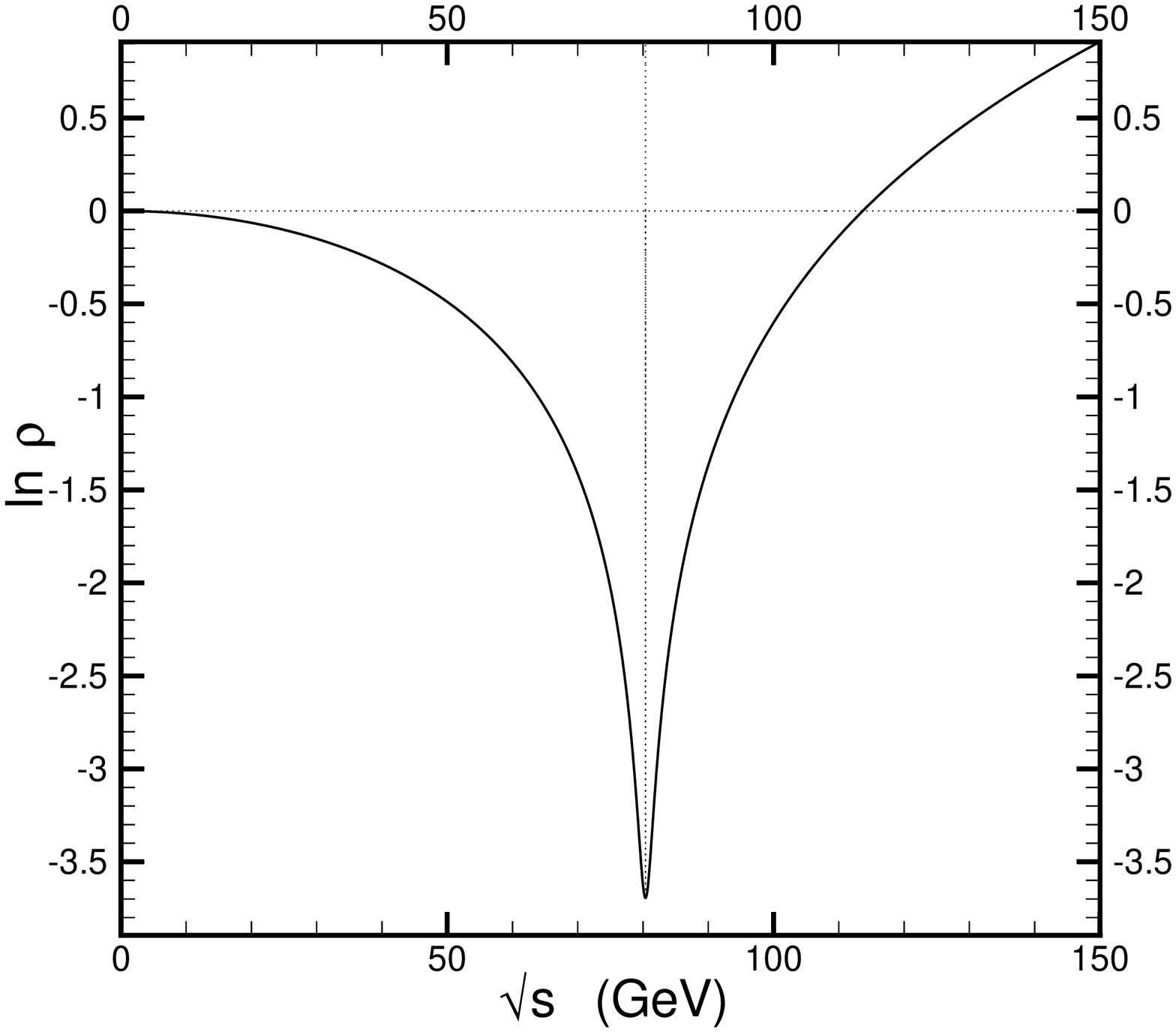,width=9.5cm}$$
\nobreak
Fig.4 The function $\ln\rho(s)$ over a large range of $\sqrt{s}$
values, for $m_1=80.4\gev$ and $\Gamma_1=2\gev$ 
(see Eq.~(\ref{eq:rho})). The minimum occurs at $\sqrt{s}=m_2$.

$$\epsfig{figure=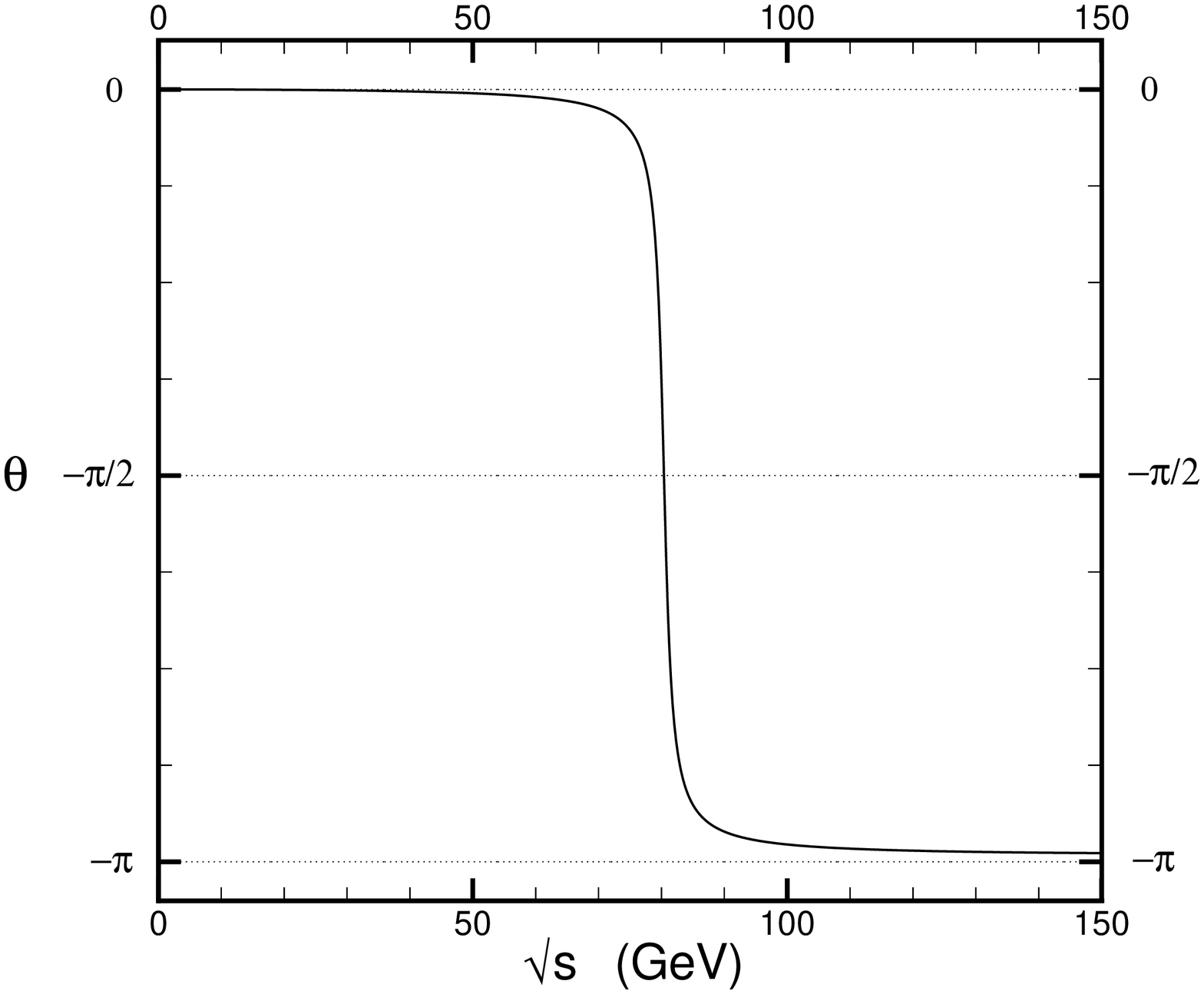,width=9.5cm}$$
\nobreak
Fig.5 The function $\theta(s)$ for $m_1=80.4\gev$ and
$\Gamma_1=2\gev$ (see Eq.~(\ref{eq:rhosintheta})). The value 
$-\pi/2$ is attained at $\sqrt{s}=m_1$.

$$\epsfig{figure=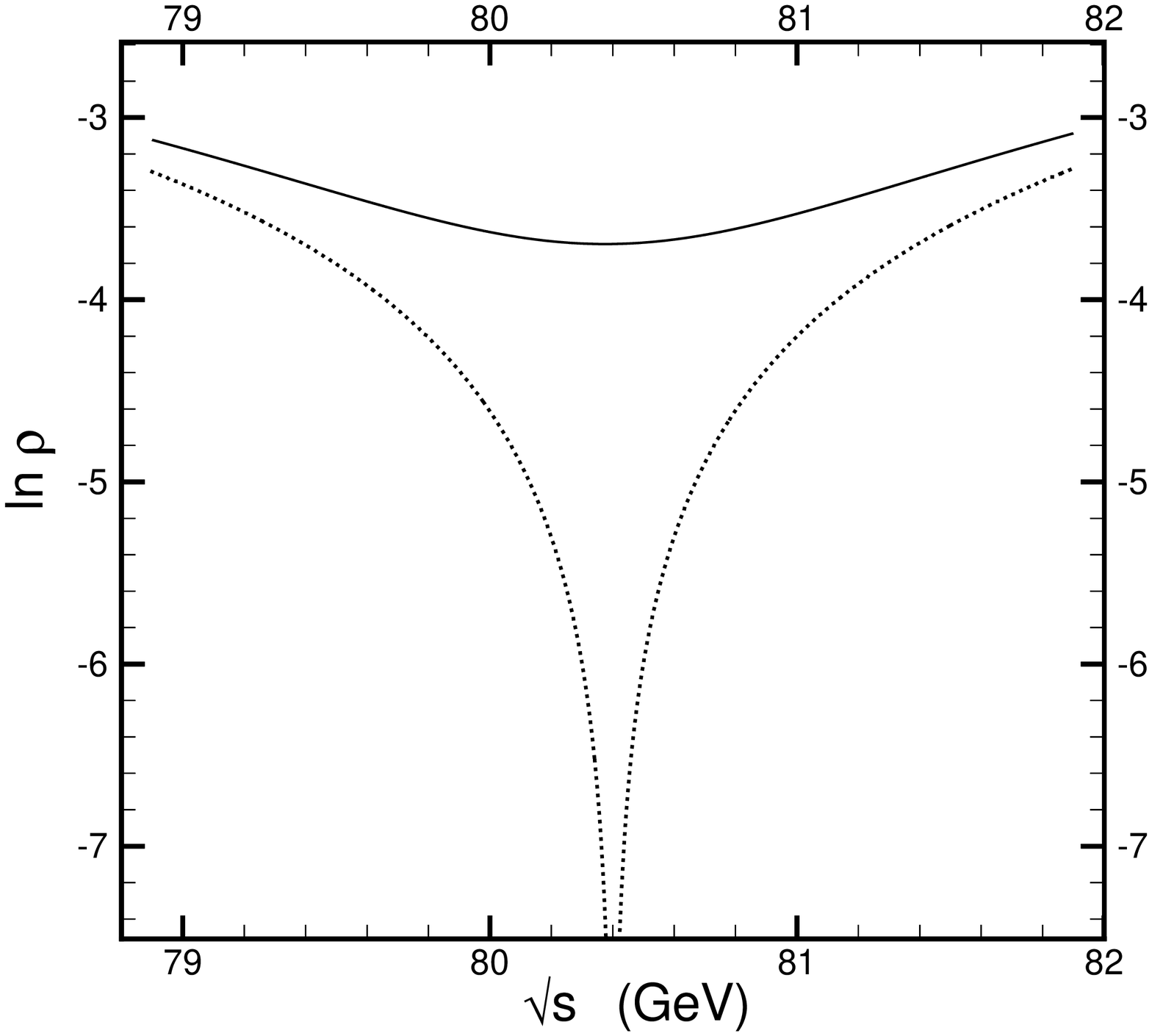,width=9.5cm}$$
\nobreak
Fig.6 Comparison of $\ln \rho(s)$ (solid line) with its zero-width 
approximation $\ln |1-s/m_1^2|$ (dotted line) over the resonance
region ($m_1=80.4\gev$, $\Gamma_1=2\gev$).

$$\epsfig{figure=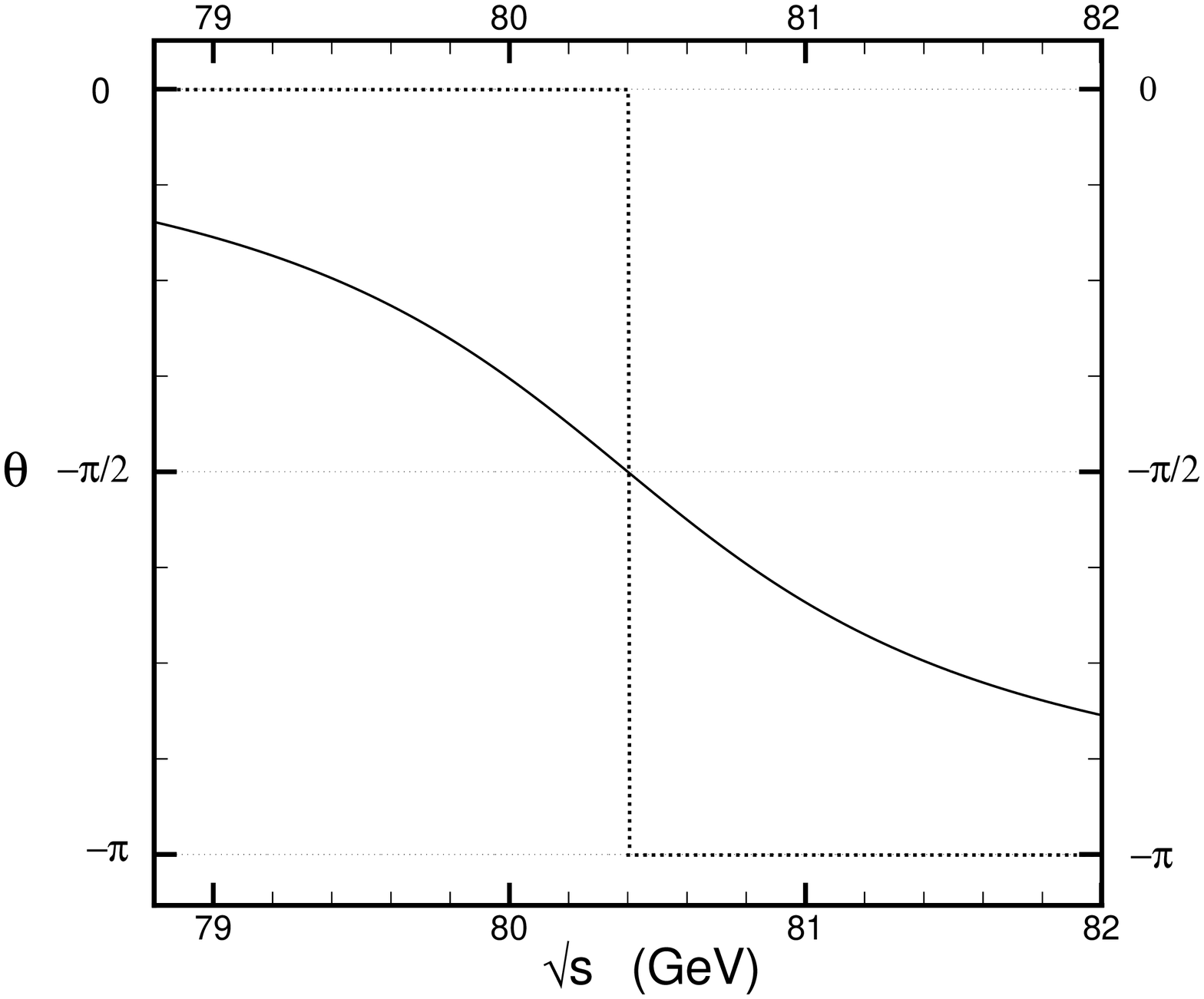,width=9.5cm}$$
\nobreak
Fig.7 Comparison of $\theta(s)$ (solid line) with the step function 
approximation (dotted line) over the resonance region
($m_1=80.4\gev$, $\Gamma_1=2\gev$).

\end{document}